\documentclass[showpacs,prl,floatfix,superscriptaddress,twocolumn]{revtex4}
\usepackage{graphicx}
\usepackage{amsmath,amsfonts}
\newcommand{\figref}[1]{Fig.~\ref{#1}}
\newcommand{\eqnref}[1]{Eq.~\eqref{#1}}

\begin{document}
\title{Stimulated and spontaneous optical generation of electron spin coherence \\ in charged GaAs quantum dots}

\author{M. V. Gurudev Dutt}
\author{Jun Cheng}
\author{Bo Li}
\author{Xiaodong Xu}
\author{Xiaoqin Li}
\author{P. R. Berman}
\author{D. G. Steel}
\email{dst@umich.edu}
\affiliation{The H. M. Randall Laboratory of Physics,
            The University of Michigan,
            Ann Arbor, MI 48109}
\author{A. S. Bracker}\author{D. Gammon}
\affiliation{The Naval Research Laboratory,
            Washington D. C. 20375}
\author{Sophia E. Economou}\author{Ren-Bao Liu}\author{L. J. Sham}
\affiliation{Department of Physics,
        The University of California-San Diego,
        La Jolla, CA 92093}

\date{\today}

\begin{abstract}
We report on the coherent optical excitation of electron spin polarization in
the ground state of charged GaAs quantum dots via an intermediate charged exciton (trion) state. Coherent optical fields are used for the creation and detection of the Raman spin coherence between the spin ground states of the charged quantum dot. The measured spin decoherence  time, which is likely limited by the nature of the spin ensemble, approaches 10~ns at zero field. We also show that the  Raman spin coherence in the quantum beats is caused not only by the usual stimulated Raman interaction but also by simultaneous spontaneous radiative decay of either excited trion state to a coherent combination of the two spin states.
\end{abstract}

\pacs{78.67.Hc, 42.50.Md, 42.50.Lc, 72.25.Rb}
\maketitle

The physics of quasi-zero dimensional semiconductor nanostructures or quantum dots (QDs)~\cite{QDs}, and semiconductor electron spins, has been extensively investigated for developments in both basic science and technology~\cite{Spintronics}.  The weak interaction of the electron spin in a QD with the environment is believed to lead to a long spin decoherence time ($T_2 \sim 1-100\ \mu$s)~\cite{SpinT2,RDSousaPRB03}. Recent measurements in GaAs and In(Ga)As QDs have also shown long spin relaxation times ($T_1 \sim 1-20$~ms)~\cite{spinT1}, which ultimately limits  $T_2$. However, coherent preparation and detection of the spin states is a challenge remaining to be addressed, with recent work in silicon devices demonstrating the rapid progress being made in this area~\cite{singlespin}.

In this Letter, we report  on the coherent optical generation of  electron spin polarization in the ground state of  charged GaAs QDs via resonant excitation of an intermediate charged exciton (trion) state. Stimulated Raman excitation by coherent optical pulses of the Raman coherence between the spin  ground states of the QD is thus demonstrated. The results will further show that the spin decoherence time is $\sim$ 10 nsec at zero field and that there are two contributions to the spin coherence: an induced part arising from coherent optical coupling of the spin states through stimulated Raman excitation, and a  \textit{spontaneously generated coherence (SGC)} arising from radiative decay of the trion into the spin states.
\begin{figure}
\scalebox{0.8}[0.7]{\includegraphics{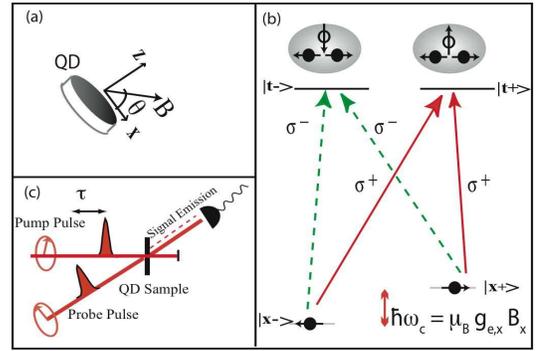}}
\caption{(a) Magnetic field ($\boldsymbol{\mathrm{B}}$) orientation with respect to QD. (b) Excitation picture for the charged QD in the Voigt geometry ($\theta = 0^\circ $), with ground states $\lvert x \pm \rangle$ denoting electron spin projections $\pm \tfrac{1}{2}$ along the $x$-axis split by $ \hbar \omega_c$. The trion states $\lvert t \pm \rangle$ are labeled by the heavy-hole angular momentum projection $\pm \tfrac{3}{2}$ along the $z$-axis. Solid (dashed) lines denote transitions excited by $\sigma^{+}(\sigma^{-})$ light. (c) Schematic of differential transmission (DT) experimental setup.  }
\label{Tlevels}
\end{figure}

 In the absence of magnetic fields, the Zeeman sublevels of the electron in the conduction-band ground state of a charged QD are assumed degenerate. The lowest optically excited state is a trion state consisting of a singlet pair of electrons and a heavy hole with its spin pointing up or down along the growth axis, designated as the $z$ axis. The  application of a magnetic field in the Voigt geometry [$\theta = 0^\circ $ in \figref{Tlevels}(a)] leads to the excitation level scheme shown in \figref{Tlevels} (b) \cite{ImamogluFP00}. This scheme is equivalent to the three-level $\Lambda$-systems used in demonstration of various quantum-optical phenomena, such as electromagnetically induced transparency~\cite{eit}, lasing without inversion~\cite{lwi}, and slowing  and freezing of light~\cite{slowlight}, which rely on the existence of a ground state Raman coherence~\cite{BinderPRL}.
 
The physical origin of the stimulated Raman coherence  can be understood in terms of the optical orientation of the spin in the growth (light propagation) direction. The spin states quantized in the $z$-direction are superpositions of the spin states quantized in the magnetic field direction [$|z\pm\rangle\equiv \left(|x+\rangle\pm|x-\rangle\right)/\sqrt{2}$], and thus the spin vector oriented along the $z$-direction represents coherence between the spin eigenstates $| x \pm \rangle$ in the magnetic field. For simplicity, we assume that before the application of the pump pulse, the system is unpolarized. In the short-pulse limit, the $\sigma^-$ pulse excites only the spin down state $|z-\rangle$ to the trion state $|t-\rangle$, leaving the electron polarized in the spin up $z$ direction. In the transverse magnetic field, the induced spin polarization (or Raman coherence) precesses and thus the optically generated population of the spin state $|z-\rangle$ oscillates as
\begin{equation}
\rho^R_{z-,z-}(\tau) =  - {\rho_{t-,t-}}\frac{1+ e^{-\gamma_s
\tau}\cos\left({\omega_c \tau} - \phi \right)}{2}, \label{eq:SRpart}
\end{equation}
where $\rho_{t-,t-}$, proportional to the intensity of the pump pulse, denotes the population of the trion state immediately after the pump pulse, and  $\omega_c, \gamma_s$ are the Larmor frequency and spin decoherence rate respectively. For the case of stimulated Raman quantum beats (QBs) given by \eqnref{eq:SRpart}, the phase $\phi = 0$.

The stimulated Raman processes which yield the QBs in \eqnref{eq:SRpart} can also be illustrated with a perturbation sequence using the density matrix operator $\rho$  for the four-level scheme shown in \figref{Tlevels}. The pump pulse ($\boldsymbol{\mathrm{E}_{1}}  (t) $) coherently excites the Raman coherence ($\rho_{\, x+,x-}$) to second order in the field, and the probe pulse ($\boldsymbol{\mathrm{E}_2} (t - \tau)$) converts this to a nonlinear polarization ($ \propto \rho_{\, t-,x-}$) which co-propagates with the probe field~\cite{BeterovPQE},
\[
\rho_{\,{x-,x-}} \xrightarrow{\mathbf{E}_{1}(\sigma^{-})}
      \rho_{\,{t-,x-}}\xrightarrow{\mathbf{E}_{1}^{*}(\sigma^{-})}
\rho_{\,{x+,x-}}
\xrightarrow{\mathbf{E}_{2}(\sigma^{\mp})}
\left\{ \begin{array}{c}
        \rho_{\,{t-,x-}}\\
        \rho_{\,{t+,x-}}
    \end{array} \right.
\]
The two upper levels are necessary in order to correctly account for the polarization dependence, but the QBs are present in the three-level $\Lambda$-system.  Within the sample, the four-wave mixing (FWM) signal field arises from the third order nonlinear optical polarization (\figref{Tlevels}(c)). The differential transmission (DT) signal represents the homodyne detection of the FWM response propagating collinearly with the reference probe field~\cite{note1,KylePRL,AnthonyPRL}. 
 In our experiments, the pump and probe laser fields can have either parallel (PCP) or orthogonal circular polarization (OCP), and the DT signal was obtained using balanced phase-sensitive detection. 

The above picture of spin QBs arising from stimulated excitation of the Raman coherence has been extremely successful~\cite{KylePRL,AnthonyPRL,ElainePRB,TartakovskiiPRL,GuptaPRB99}. However, it has also been anticipated that the spontaneous emission from the trion state can result in a coherent combination of the two spin states, known as spontaneously generated coherence (SGC)~\cite{sgc,spemcanc}. Spontaneous emission is commonly considered to destroy coherence, but when (i) the Zeeman splitting $\hbar \omega_c$ is comparable to or smaller than the trion
decay rate $2\Gamma_t$, and (ii) the transition dipole moments from the trion state to the two spin states are non-orthogonal, different transitions can couple to the same modes of the electromagnetic vacuum. For instance, the transitions between $|t-\rangle$  and $|x\pm\rangle$   can couple to the same vacuum mode with polarization $\sigma^-$, thus creating coherence between the final states ($|x\pm\rangle$) of the spontaneous emission process. In the optical orientation picture, the trion state $|t-\rangle$ will relax back to the  state $|z-\rangle$ by spontaneously emitting a $\sigma^-$ polarized photon, thus giving rise to coherence between the energy eigenstates $|x+\rangle$ and $|x-\rangle$. Although there have been numerous theoretical studies previously~\cite{sgc,spemcanc}, there had been no experimental observation of an excited-state population decaying to a ground state Raman coherence. From the detailed analysis to be presented later, a portion of the Raman coherence caused by SGC is $\pi/2$ out of phase and interferes destructively with the stimulated Raman coherence, giving rise to a non-zero phase $\phi$ in \eqnref{eq:SRpart}. 

 For our work, the sample consisted of interface fluctuation GaAs QDs, formed by growth interrupts at the interface of a narrow (4.2 nm) GaAs quantum well, which were remotely doped with electrons. Magneto-photoluminescence measurements on single charged QDs showed a small electron $g$-factor~\cite{TischlerPRB} that implies a maximum splitting $\hbar \omega_{c} \sim 80 \;\mu$eV at the highest fields reached in our experiments. The etched sample was mounted in a superconducting magnetic cryostat  held at  4.8 K.  The pump  and probe  pulses were obtained from a mode-locked Ti:Sapphire laser, with a shaped pulse bandwidth (FWHM = 0.84 meV)  that exceeds the splitting between the electron spin states. 

 \begin{figure}
\scalebox{0.9}[0.75]{\includegraphics[width=3.4in]{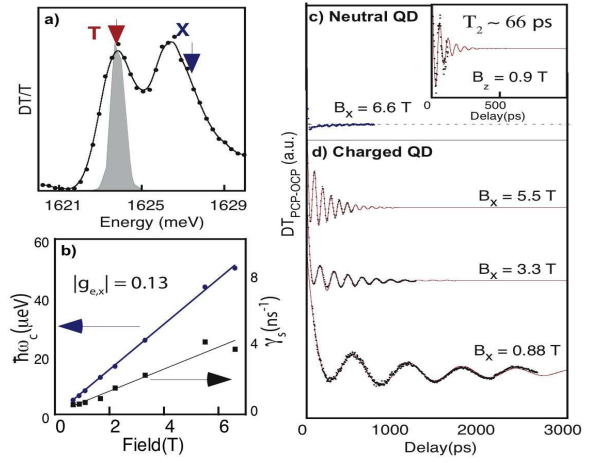}}
\caption{(a) DT spectrum with the pulse delay fixed at +10 ps. The shaded region is the pulse spectrum, and the arrows at T (X) label the trion (exciton) resonances. (b) Variation of the decay rate and the oscillation frequency with the field. We extract the $g$-factor to be $\lvert g_{e,x}  \rvert = 0.13 $. (c)  DT signal at X for B$_x$ = 6.6 T showing no oscillations. Inset: Single QD exciton QBs obtained in the Faraday geometry with B$_z$ = 0.9 T~\cite{ElainePRB}. (d) Spin QBs in DT signal as a function of in-plane magnetic field B$_x$, obtained when the laser pulse is tuned to selectively excite the T resonance. Data shown in (c) and (d) are the differences between PCP and OCP signals. Solid lines show fits to the data using \eqnref{eq:DTdiff}. }
\label{F2}
\end{figure}
\figref{F2}(a) shows the nonlinear optical spectrum with two peaks corresponding to trions (T) and excitons (X) trapped in charged and neutral QDs, respectively~\cite{TischlerPRB}. In \figref{F2}(c) [(d)] we plot the difference between DT signal obtained for the PCP and OCP configurations, with the laser tuned to resonantly excite neutral [charged] QDs at X [T]. The data in \figref{F2}(d) was fitted using the equation,
\begin{equation}
DT_{PCP-OCP} =  A_{1} e^{- 2\Gamma_t \tau} + A_{2} e^{- \gamma_s \tau} \cos(\omega_c \tau - \phi)
\label{eq:DTdiff}
\end{equation}
which was obtained from the theory~\cite{Sophia}, and the resulting values for $\hbar \omega_c$ and  $\gamma_s$ are plotted in \figref{F2}(b) as a function of the field. The oscillation frequency ($\omega_c$) depends on the splitting between the states and allows us to find  $\lvert g_{e,x} \rvert = 0.13$ in good agreement with~\cite{TischlerPRB}. The theory shows that the first term in \eqnref{eq:DTdiff} arises from the incoherent contribution due to the Pauli blocking effect, and decays with the trion recombination rate $2 \Gamma_t \sim (84 \, \text{ps})^{-1}$. The second term, describing the beats, arises from the coherent contribution ($\rho_{x+,x-}$) or the net precessing spin polarization in the $z$ direction, and decays with the spin decoherence rate $\gamma_s$. Hole spin relaxation between the two trion states was accounted for and affects the Pauli blocking term only, but not the QBs.

The spin dephasing time $T_{2}^{*}(=1/\gamma_s) \sim 10$~ns was obtained from the zero-field intercept in \figref{F2}(b), which is much smaller than the $T_2 $ predicted by theory~\cite{SpinT2,RDSousaPRB03}.   Electrons in different dots or in the same dot at
different times experience different random orientations of the hyperfine nuclear fields. If in each dot the electron interacts with $N \sim 10^6$ nuclei,  the characteristic timescale due to the spatial fluctuations in the hyperfine interaction is $T_{N} \approx \hbar \sqrt{N}/ A = 6$ ns ~\cite{SpinT2}, where $A \approx 90 \,\mu$eV is the hyperfine constant for GaAs, comparable to the decay times obtained here. Our ensemble measurements are consistent with a previous Hanle study of a single dot in the same system~\cite{AllanPRL05}. The linear change in $\gamma_s$ with the magnetic field is commonly attributed to a Gaussian distribution of $g$-factors~\cite{GuptaPRB99}, from which we obtain a width $\Delta g / g \sim 8 \% $.

In contrast to the T resonance, tuning the laser to the X resonance results in vanishing of the beats as seen in \figref{F2}(c). As discussed in Refs.~\cite{AnthonyPRL,TartakovskiiPRL,ElainePRB}, the exciton behaves as a three-level V-system. The fast exciton recombination time ($\sim$ 50-100 ps~\cite{NicoPRL98}) and the small Zeeman splitting in the Voigt geometry results in overdamping of the exciton Raman coherence.  For comparison, the inset to \figref{F2}(c) shows DT QBs of magneto-excitons in a single neutral QD obtained in the Faraday geometry ($\theta = 90^\circ$), where the splitting is enhanced by the large $g$-factor. Ref.~\cite{ElainePRB} reports that the exciton Raman decoherence time ($T_2 \sim$ 66 ps) is limited by recombination.
 
 \begin{figure}
\scalebox{0.85}[0.9]{\includegraphics{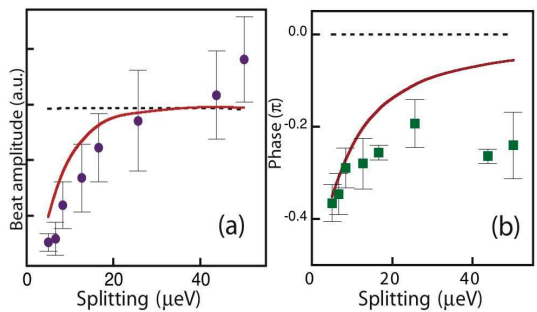}}
\caption{(a) and (b) show the changes in the amplitude (A$_2$) and phase ($\phi$) of the oscillations as a function of splitting. Solid (dashed) lines denote theoretical predictions for these parameters with (without) the effects of SGC. In numerical calculations,
the optical pulses are assumed Gaussian; i.e. $\mathcal{I}(\omega) = \exp(- \omega^{2}/2 \sigma^{2})$ ($\sigma = 0.35 \text{ meV}$).}
\label{F3}
\end{figure}
When there are no effects due to SGC, and in the limit where the pulse bandwidth is much greater than the Zeeman splitting, the amplitude and phase shift of the spin Raman coherence generated by the pump pulse should be independent of the Zeeman splitting. This is verified by the theory, as shown in the dashed lines of \figref{F3}. However, the data (solid symbols) in \figref{F3}(a) and (b) demonstrates a clear dependence of the beat amplitude (A$_2$) and phase ($\phi$) on the Zeeman splitting $\hbar \omega_c$  which has not been reported in earlier experiments~\cite{KylePRL,AnthonyPRL,ElainePRB,TartakovskiiPRL,GuptaPRB99}. We verified that the dependence is not due to shifts in the trion resonance, or changes in the oscillator strength with magnetic field. As elaborated below, the data in \figref{F3} provides experimental confirmation for SGC. 

Specifically, the  spin polarization or coherence spontaneously generated in the time interval $(t, t+dt)$ for $t < \tau$ after the pump pulse can be determined by the population generated in $|z-\rangle$~\cite{Sophia,ShabaevPRB03},
\begin{equation}
d\rho^{\text{SGC}}_{z-,z-} =  \Gamma_t  {\rho_{t-,t-}} e^{-2\Gamma_t t}
[1+e^{-\gamma_s (\tau-t)}\cos \left({\omega_c( \tau-t)}
\right)] \,dt.
\end{equation}
Thus, the spin coherence accumulated from $t=0$ to $t=\tau$ via the trion recombination process is
dependent on the Zeeman splitting. Combining the contributions from the pump-excited Raman coherence in \eqnref{eq:SRpart} and from SGC yields spin beats with amplitude and phase shift given, in the short-pulse limit,  by
\begin{eqnarray}
{A}_2 &\approx&
\sqrt{\frac{{\gamma_s}^2+\omega^2_{c}}{(2\Gamma_{t}-{\gamma_s})^2+\omega^2_{c}}},
\\
\phi&\approx&
-\arctan\left(\frac{2\Gamma_{t}-\gamma_s}{\omega_{c}}\right)
-\arctan\left(\frac{\gamma_s}{\omega_{c}}\right). \label{old16}
\end{eqnarray}
In the weak magnetic field limit, the trion decay is much faster than the spin precession, and so the SGC cancels the conventional Raman coherence. In the strong magnetic field limit, the rapid spin precession averages the SGC to zero. This explains the observed field dependence of the spin beats in \figref{F3}. As the Zeeman splitting increases from zero to much larger than the radiative decay rate,
the beat amplitude increases until it saturates at the value calculated without the SGC effect, and the phase shift increases from close to $-\pi/2$ to zero.

In the master equation for the $\Lambda$ system, derived in the Markov-Born approximation, the SGC effect appears as,
\begin{equation}
\dot{\rho}_{\,{x+,x-}} \rvert_{\text{SGC}} =  -\Gamma_{t}
\rho_{\,{t-,t-}}.
\end{equation}
A full numerical solution of the master equation, taking into account the finite pulse width, is performed at each magnetic field value for the time delay dependence of the DT signal, with the input parameters taken from experiment for the Zeeman splitting, spin dephasing rate, and the trion decay rate~\cite{Sophia}. Thus, the amplitude and phase of the spin beats computed as a function of the Zeeman splitting are shown as solid lines in \figref{F3}, similar to the results obtained in the short-pulse limit. The agreement of the SGC theory with experimental data is good. The deviation of the theory and experiment as a function of magnetic field compared to the expected behavior in the absence of SGC is evident.

Physically, as discussed above, the same mode of the vacuum field simultaneously couples the trion state to the two spin states and gives rise to SGC in our experiments. In atomic systems, ground state Raman coherence induced by population decay from an excited state has been difficult to observe due to the conditions (i) and (ii) that must be simultaneously met. However, recently an entangled state between a single ion and two modes of the radiation field was observed~\cite{monroe}, an effect which can be interpreted as complementary to SGC~\cite{Sophia}. The observation of SGC in QDs is favored not only by the relatively large radiative decay rate as compared to the spin precession rate, but also by the ability to tailor the quantum states in the artificial atoms by magnetic field and by doping.  

The results of this work bear directly on proposals for using optically driven charged QDs for quantum computing (QC)~\cite{ImamogluFP00,opticalSpinQC}. In this regard, significant progress has already been made in demonstrating many of the key DiVincenzo requirements for QC~\cite{DiVincenzoFP00} with excitonic optical Bloch vector qubits in neutral QDs, such as single qubit rotations~\cite{SingleXRabi}, quantum entanglement~\cite{GangSci00} and conditional quantum logic gates~\cite{ElaineSci03}. These results illustrate the close analogy between artificial atoms and QDs~\cite{QDs}, but the short decoherence time of the exciton (0.1 -- 1 ns~\cite{NicoPRL98,BorriPRL01}) constrains quantum error correction schemes. In the present work, we have demonstrated that a coherent optical field can create a coherent superposition of electron spin states in QDs. While the measured decoherence time has a limiting value of 10 ns, much longer than the exciton decoherence time, the lifetime is shorter than anticipated due to ensemble averaging, as discussed previously. Coherent spectroscopy measurements on a single dot could eliminate some aspects of this averaging. Single qubit rotations are a logical next step remaining to be pursued, for example through off-resonant Raman processes as proposed in Ref.~\cite{chen}. Modifications to the selection rules, caused by the small heavy-light-hole mixing ($\sim 0.06$), will merely result in slight changes to the axis and angle of the Raman induced spin qubit rotation, which can be compensated optically. \begin{acknowledgments}
This work was supported in part by the U.S. ARO, NSA, ARDA, AFOSR, ONR and the NSF. The authors would also like to thank A. Bragas, J. Bao and  R. Merlin for helpful discussions.
\end{acknowledgments}


\begin{thebibliography}{10}
\bibitem{note1}
In the absence of mutual pump/probe coherence, DT is simply the difference between the probe absorption with and without the pump excitation. See also Refs.~\cite{KylePRL,AnthonyPRL}.
\end{thebibliography}
\end{document}